\documentstyle[prl,aps]{revtex}
\begin{document}
\draft
\title{A $n$-qubit controlled phase gate with superconducting quantum interference
devices coupled to a resonator}
\author{Chui-Ping Yang and Siyuan Han}
\address{Department of Physics and Astronomy, University of Kansas, Lawrence,\\
Kansas 66045}
\maketitle

\begin{abstract}
We present a way to realize a $n$-qubit controlled phase gate with
superconducting quantum interference devices (SQUIDs) by coupling them to a
superconducting resonator. In this proposal, the two logical states of a
qubit are represented by the two lowest levels of a SQUID. An intermediate
level of each SQUID is utilized to facilitate coherent control and
manipulation of quantum states of the qubits. It is interesting to note that
a $n$-qubit controlled phase gate can be achieved with $n$ SQUIDs by
successively applying a $\pi /2$ Jaynes-Cummings pulse to each of the $n-1$
control SQUIDs before and after a $\pi$ Jaynes-Cummings pulse on the target
SQUID.
\end{abstract}

\date{\today }


\begin{center}
{\bf I. INTRODUCTION AND MOTIVATION}
\end{center}

Quantum computing has attracted much interest since it was clear that
quantum computers are in principle able to solve hard computational problems
much more efficiently than classical computers [1-3]. In the past decade,
various physical systems have been considered for building up quantum
information processors. Among them, cavity QED analogs with solid state
systems are particularly appealing. Theoretically, it was predicted earlier
that the strong coupling limit, which is difficult to achieve with atoms in
a microwave cavity, can readily be realized with superconducting charge
qubits [4,5], superconducting flux qubits [6], or semiconducting quantum
dots [7]. Recently, the strong coupling cavity QED has been experimentally
demonstrated with superconducting charge qubits and flux qubits [8,9] and
semiconductor quantum dots embedded in a microcavity [10-12]. The results of
these experiments make solid state qubit cavity QED a very attractive
approach to quantum information process.

It is known that a quantum computing network can be decomposed into
two-qubit gates and one-qubit rotations [13]. So far, a large number of
theoretical proposals for realizing two-qubit gates have been presented with
many physical systems. Moreover, two-qubit CNOT gates or controlled phase
gates have been experimentally demonstrated in cavity QED [14], ion trap
[15], NMR [16], quantum dots [17], and superconducting charge qubits [18].
On the other hand, research on quantum computing has recently moved towards
the physical realization of multiqubit quantum gates. Several schemes for
realizing three-qubit Toffoli gates have been proposed with neutral atoms in
an optical lattice [19] or hybrid atom-photon qubits via cavity QED [20]. In
addition, experimental realization of a controlled phase gate in a three
qubit NMR quantum system has been reported recently [21].

In this paper, our goal is to present a way to realize a $n$-qubit
controlled phase gate with superconducting quantum interference devices
(SQUIDs), coupled to a superconducting resonator. In this proposal, the two
lowest levels $\left| 0\right\rangle $ and $\left| 1\right\rangle $ of each
SQUID represent the two logical states of a qubit while a higher-energy
level $\left| 2\right\rangle $ of each SQUID is used to facilitate coherent
control and manipulation of quantum states of the qubits (Fig. 1). The
method presented here essentially operates by having the resonator to be:
(a) resonant with the $\left| 0\right\rangle \leftrightarrow \left|
2\right\rangle $ transition of each SQUID sequentially, while (b) largely
detuned from the $\left| 1\right\rangle \leftrightarrow \left|
2\right\rangle $ transition and the $\left| 0\right\rangle \leftrightarrow
\left| 1\right\rangle $ transition of each SQUID during the gate operation.
We find that a $n$-qubit controlled phase gate can be achieved with $n$
SQUIDs, by successively applying a $\pi /2$ Jaynes-Cummings pulse to each of
the $n-1$ control SQUIDs before and after a $\pi $ Jaynes-Cummings pulse on
the remaining SQUID (the target qubit).

As shown below, this scheme has the following advantages: (i) No tunneling
between the qubit levels $\left| 0\right\rangle $ and $\left| 1\right\rangle 
$ is required so that the storage time of each qubit can be made very long;
(ii) No measurement on SQUIDs or photons is needed, therefore the operation
is simplified; (iii) No auxiliary SQUIDs are needed, thus saving precious
hardware resources; and (iv) Coupling constants of each SQUID with the
resonator could be non-identical to accommodate inevitable nonuniformity in
device parameters.

The motivation for this work is threefold: (i) SQUIDs have recently
attracted much attention in quantum information community. Reasons for this
are that SQUIDs are relatively easy to scale up and have been demonstrated
to have relatively long decoherence times among solid state qubits [9,22-27]
and therefore have been considered as a promising candidate for building up
superconducting quantum computers/information processors [28-37]. (ii) In
the present proposal, the strong coupling between a SQUID and the resonator
is analogous to atomic cavity quantum electrodynamics (QED). Although many
schemes for realizing two-qubit quantum gates with a variety of physical
systems using cavity QED [5,32-34,38-45] or trapped ions [46-49] have been
proposed, how to realize a multiqubit controlled quantum gate based on
cavity QED or trapped ions has not been thoroughly investigated. (iii) As is
well known, multiqubit controlled quantum gates are of great importance to
constructing quantum computational networks, realizing quantum error
correction protocols, and implementing quantum algorithms.

This paper is organized as follows. In Sec. II, we review the basic theory
of a SQUID coupled to a single-mode resonator or driven by a classical
microwave pulse. In Sec. III, we show how to realize a three-qubit
controlled phase gate with three SQUIDs in a resonator. We then discuss how
the method can be generalized to implement a $n$-qubit controlled phase gate
with $n$ SQUIDs in a resonator. In Sec. IV, we give a discussion on the
experimental issues. A concluding summary is given in Sec. V.

\begin{center}
{\bf II. BASIC THEORY}
\end{center}

The SQUIDs considered throughout this paper are rf SQUIDs each consisting of
a Josephson tunnel junction enclosed by a superconducting loop (the size of
an rf SQUID is on the order of 10 $\mu $m $-$ 100 $\mu $m). The Hamiltonian
for an rf SQUID (with junction capacitance $C$ and loop inductance $L$) has
the usual form [25] 
\begin{equation}
H_s=\frac{Q^2}{2C}+\frac{\left( \Phi -\Phi _x\right) ^2}{2L}-E_J\cos \left(
2\pi \frac \Phi {\Phi _0}\right) ,
\end{equation}
where $\Phi $, the magnetic flux threading the ring, and $Q$, the total
charge on the capacitor, are the conjugate variables of the system (with the
commutation relation $\left[ \Phi ,Q\right] =i\hbar $), $\Phi _x$ is the
static (or quasistatic) external magnetic flux applied to the ring, and $E_J$
$\equiv I_c\Phi _0/2\pi $ is the Josephson coupling energy ($I_c$ is the
critical current of the junction and $\Phi _0=h/2e$ is the flux quantum).

\begin{center}
{\bf A. SQUID coupled to a single-mode resonator}
\end{center}

Consider a SQUID coupled to a single-mode resonator. The SQUID is biased
properly to have the $\Lambda $-type three lowest levels, which are denoted
by $\left| 0\right\rangle ,$ $\left| 1\right\rangle $ and $\left|
2\right\rangle $, respectively (Fig. 1). Suppose that the coupling of $%
\left| 0\right\rangle ,\left| 1\right\rangle $ and $\left| 2\right\rangle $
with other levels of the SQUID via the resonator is negligible, which can be
readily achieved by adjusting the level spacings of the SQUID. It can be
shown that when the resonator is resonant with the $\left| 0\right\rangle
\leftrightarrow \left| 2\right\rangle $ transition while far-off resonant
with the $\left| 1\right\rangle \leftrightarrow \left| 2\right\rangle $
transition and the $\left| 0\right\rangle \leftrightarrow \left|
1\right\rangle $ transition of the SQUID, the interaction Hamiltonian in the
interaction picture, after making the rotating-wave approximation, can be
written as [32] 
\begin{equation}
H_I=\hbar \left( ga^{+}\left| 0\right\rangle \left\langle 2\right| +\text{%
h.c.}\right) .
\end{equation}
Here, $a^{+}$ and $a$ are the creation and annihilation operators of the
resonator mode with frequency $\nu _r,$ and $g$ is the coupling constant
between the resonator mode and the $\left| 0\right\rangle \leftrightarrow
\left| 2\right\rangle $ transition of the SQUID. For a superconducting
one-dimensional standing-wave resonator, the expression of $g$ is given by 
\begin{equation}
\hbar g\left( x\right) =\frac{M_{sr}}L\sqrt{\frac{h\nu _r}{L_0l}}%
\left\langle 0\right| \Phi \left| 2\right\rangle \sin \left( \frac{2\pi }%
\lambda x\right) ,
\end{equation}
where $M_{sr}$ is the mutual inductance between the SQUID and the resonator, 
$L_0$ is the inductance per unit length of the resonator, $l$ is the length
of the resonator, $\nu _r$ is the frequency of the resonator mode with
wavelength $\lambda ,$ and $x$ is the center of the SQUID in the resonator.

The Hamiltonian (2) actually has the same form as the Jaynes-Cummings
Hamiltonian of a 2-level system resonant with a single-mode cavity field (a
Hamiltonian well known in quantum optics). In the case when the resonator is
initially in the photon-number state $\left| n\right\rangle $, the time
evolution of the states of the system, governed by the Hamiltonian (2), is
described by 
\begin{eqnarray}
\left| 0\right\rangle \left| n\right\rangle &\rightarrow &\cos \sqrt{n}%
gt\left| 0\right\rangle \left| n\right\rangle -i\sin \sqrt{n}gt\left|
2\right\rangle \left| n-1\right\rangle ,  \nonumber \\
\left| 2\right\rangle \left| n\right\rangle &\rightarrow &-i\sin \sqrt{n+1}%
gt\left| 0\right\rangle \left| n+1\right\rangle +\cos \sqrt{n+1}gt\left|
2\right\rangle \left| n\right\rangle .
\end{eqnarray}
Note that the coupling strength $g$ may vary with different SQUIDs due to
non-uniform device parameters and/or non-exact placement of SQUIDs in the
cavity. Therefore, hereafter we replace $g$ by $g_1,$ $g_2,...,$ and $g_n$
for SQUIDs $1,2,...,$ and $n,$ respectively.{\bf \ }

\begin{center}
{\bf B. SQUID driven by a classical microwave pulse}
\end{center}

Now let us consider a SQUID driven by a classical microwave pulse with the
magnetic component ${\bf B}_{\mu w}({\bf r,t})={\bf B}_{\mu w}({\bf r})$ cos 
$\left( 2\pi \nu _{\mu w}t+\phi \right) $. Here, ${\bf B}_{\mu w}({\bf r),}$ 
$\nu _{\mu w}$, and $\phi $ are the magnetic field amplitude, frequency, and
phase of the microwave pulse. It can be shown that if the microwave pulse is
resonant with the $\left| 1\right\rangle \leftrightarrow \left|
2\right\rangle $ transition but far-off resonant with the $\left|
0\right\rangle \leftrightarrow \left| 2\right\rangle $ transition and the $%
\left| 0\right\rangle \leftrightarrow \left| 1\right\rangle $ transition of
the SQUID, then the interaction Hamiltonian in the interaction picture is
given by 
\begin{equation}
H_I=\frac \hbar 2\left( \Omega _{12}e^{i\phi }\left| 1\right\rangle
\left\langle 2\right| +\text{h.c.}\right) ,
\end{equation}
where $\Omega _{12}$ is the Rabi frequency of the pulse, which takes the
following form [32] 
\begin{equation}
\Omega _{12}\left( t\right) =\frac 1{L\hbar }\left\langle 1\right| \Phi
\left| 2\right\rangle \int_S{\bf B}_{\mu w}({\bf r})\cdot d{\bf S.}
\end{equation}
From the Hamiltonian (5), it is straightforward to see that a pulse of
duration $t$ results in the following rotation 
\begin{eqnarray}
\left| 1\right\rangle &\rightarrow &\cos \frac{\Omega _{12}}2t\left|
1\right\rangle -ie^{-i\phi }\sin \frac{\Omega _{12}}2t\left| 2\right\rangle ,
\nonumber \\
\left| 2\right\rangle &\rightarrow &-ie^{i\phi }\sin \frac{\Omega _{12}}2%
t\left| 1\right\rangle +\cos \frac{\Omega _{12}}2t\left| 2\right\rangle .
\end{eqnarray}

\begin{center}
{\bf III. MULTIQUBIT CONTROLLED PHASE GATE WITH SQUIDS}
\end{center}

In this section, for clarity, we will first give an explicit description on
how to realize a three-qubit controlled phase gate with three SQUIDs coupled
to a microwave resonator. We will then discuss how to extend the method to
obtain a $n$-qubit controlled phase gate with a larger number $n$.

\begin{center}
{\bf A. Three-qubit controlled phase gate}
\end{center}

For three qubits, there are a total number of eight ($2^3$) computational
basis states, denoted by $\left| 000\right\rangle ,\left| 001\right\rangle
,...,\left| 111\right\rangle ,$ respectively. A three-qubit controlled phase
gate is described by 
\begin{eqnarray}
\left| 000\right\rangle &\rightarrow &\left| 000\right\rangle ,\;\left|
100\right\rangle \rightarrow \left| 100\right\rangle ,  \nonumber \\
\left| 001\right\rangle &\rightarrow &\left| 001\right\rangle ,\;\left|
101\right\rangle \rightarrow \left| 101\right\rangle ,  \nonumber \\
\left| 010\right\rangle &\rightarrow &\left| 010\right\rangle ,\;\left|
110\right\rangle \rightarrow \left| 110\right\rangle ,  \nonumber \\
\left| 011\right\rangle &\rightarrow &\left| 011\right\rangle ,\;\left|
111\right\rangle \rightarrow -\left| 111\right\rangle ,
\end{eqnarray}
which implies that if and only if the two control qubits (the first two
qubits) are in the state $\left| 1\right\rangle $, a phase flip happens to
the state $\left| 1\right\rangle $ of the target qubit (the last qubit) and
nothing happens otherwise.

To realize this gate, consider SQUIDs ($1,2,3$) each having the $\Lambda $%
-type level configuration as depicted in Fig. 1. The transition between any
two levels for each SQUID is initially far-off resonant with the resonator
(e.g., via prior adjustment of the level spacings) and the cavity mode is
initially in the vacuum state $\left| 0\right\rangle _c.$

The operations for realizing the three-qubit controlled phase gate are
listed below:

Step (i): Apply a $\pi $ microwave pulse ($\Omega _{12}\tau _{\mu w}=\pi ,$
where $\tau _{\mu w}$ is the pulse duration) with $\phi =-\pi /2$ to SQUID $%
1 $ [Fig. 2(a)]. The pulse is resonant with the $\left| 1\right\rangle
\leftrightarrow \left| 2\right\rangle $ transition of SQUID $1.$ After the
pulse, the transformation $\left| 1\right\rangle \rightarrow \left|
2\right\rangle $ of SQUID $1$ is obtained.

Step (ii): Bring the $\left| 0\right\rangle \leftrightarrow \left|
2\right\rangle $ transition of SQUID $1$ to resonance with the resonator for
an interaction time $\tau _1=\pi /\left( 2g_1\right) $ [Fig. 2(b)],
resulting in $\left| 2\right\rangle _1\left| 0\right\rangle _c\rightarrow
-i\left| 0\right\rangle _1\left| 1\right\rangle _c$.

Step (iii): Bring the $\left| 0\right\rangle \leftrightarrow \left|
2\right\rangle $ transition of SQUID $2$ to resonance with the resonator for
an interaction time $\tau _2=\pi /\left( 2g_2\right) $ [Fig. 2(c)]. As a
result, the states $\left| 0\right\rangle _2\left| 0\right\rangle _c,\left|
1\right\rangle _2\left| 0\right\rangle _c,$ and $\left| 1\right\rangle
_2\left| 1\right\rangle _c$ remain unchanged, while the state $\left|
0\right\rangle _2\left| 1\right\rangle _c$ changes to $-i\left|
2\right\rangle _2\left| 0\right\rangle _c.$

Step (iv): Bring the $\left| 0\right\rangle \leftrightarrow \left|
2\right\rangle $ transition of SQUID $3$ to resonance with the resonator for
an interaction time $\tau _3=\pi /g_3$ [Fig. 2(d)], resulting in $\left|
0\right\rangle _3\left| 1\right\rangle _c\rightarrow -\left| 0\right\rangle
_3\left| 1\right\rangle _c$ while no change for the states $\left|
0\right\rangle _3\left| 0\right\rangle _c,$ $\left| 1\right\rangle _3\left|
0\right\rangle _c$, and $\left| 1\right\rangle _3\left| 1\right\rangle _c.$

Step (v): Bring the $\left| 0\right\rangle \leftrightarrow \left|
2\right\rangle $ transition of SQUID $2$ to resonance with the resonator for
an interaction time $\tau _2=\pi /\left( 2g_2\right) $ [Fig. 2(c)]. As a
result, the states $\left| 0\right\rangle _2\left| 0\right\rangle _c,\left|
1\right\rangle _2\left| 0\right\rangle _c,$ and $\left| 1\right\rangle
_2\left| 1\right\rangle _c$ remain unchanged, while the state $\left|
2\right\rangle _2\left| 0\right\rangle _c$ becomes $-i\left| 0\right\rangle
_2\left| 1\right\rangle _c.$

Step (vi): Bring the $\left| 0\right\rangle \leftrightarrow \left|
2\right\rangle $ transition of SQUID $1$ to resonance with the resonator for
an interaction time $\tau _1=\pi /\left( 2g_1\right) $ [Fig. 2(b)],
resulting in $\left| 0\right\rangle _1\left| 1\right\rangle _c\rightarrow
-i\left| 2\right\rangle _1\left| 0\right\rangle _c.$

Step (vii): Apply a $\pi $ microwave pulse with $\phi =\pi /2$ to SQUID $1$
[Fig. 2(a)]. The pulse is resonant with the $\left| 1\right\rangle
\leftrightarrow \left| 2\right\rangle $ transition of SQUID $1$. After the
pulse, the transformation $\left| 2\right\rangle \rightarrow \left|
1\right\rangle $ of SQUID $1$ is achieved.

The states of the whole system after each step of the above operations are
summarized below: 
\[
\begin{array}{c}
\left| 100\right\rangle \left| 0\right\rangle _c \\ 
\left| 101\right\rangle \left| 0\right\rangle _c \\ 
\left| 110\right\rangle \left| 0\right\rangle _c \\ 
\left| 111\right\rangle \left| 0\right\rangle _c
\end{array}
\stackrel{\text{Step(i)}}{\longrightarrow } 
\begin{array}{c}
\left| 200\right\rangle \left| 0\right\rangle _c \\ 
\left| 201\right\rangle \left| 0\right\rangle _c \\ 
\left| 210\right\rangle \left| 0\right\rangle _c \\ 
\left| 211\right\rangle \left| 0\right\rangle _c
\end{array}
\stackrel{\text{Step(ii)}}{\longrightarrow } 
\begin{array}{c}
-i\left| 000\right\rangle \left| 1\right\rangle _c \\ 
-i\left| 001\right\rangle \left| 1\right\rangle _c \\ 
-i\left| 010\right\rangle \left| 1\right\rangle _c \\ 
-i\left| 011\right\rangle \left| 1\right\rangle _c
\end{array}
\stackrel{\text{Step(iii)}}{\longrightarrow } 
\begin{array}{c}
-\left| 020\right\rangle \left| 0\right\rangle _c \\ 
-\left| 021\right\rangle \left| 0\right\rangle _c \\ 
-i\left| 010\right\rangle \left| 1\right\rangle _c \\ 
-i\left| 011\right\rangle \left| 1\right\rangle _c
\end{array}
\]
\begin{equation}
\stackrel{\text{Step(iv)}}{\longrightarrow } 
\begin{array}{c}
-\left| 020\right\rangle \left| 0\right\rangle _c \\ 
-\left| 021\right\rangle \left| 0\right\rangle _c \\ 
i\left| 010\right\rangle \left| 1\right\rangle _c \\ 
-i\left| 011\right\rangle \left| 1\right\rangle _c
\end{array}
\stackrel{\text{Step(v)}}{\longrightarrow } 
\begin{array}{c}
i\left| 000\right\rangle \left| 1\right\rangle _c \\ 
i\left| 001\right\rangle \left| 1\right\rangle _c \\ 
i\left| 010\right\rangle \left| 1\right\rangle _c \\ 
-i\left| 011\right\rangle \left| 1\right\rangle _c
\end{array}
\stackrel{\text{Step(vi)}}{\longrightarrow } 
\begin{array}{c}
\left| 200\right\rangle \left| 0\right\rangle _c \\ 
\left| 201\right\rangle \left| 0\right\rangle _c \\ 
\left| 210\right\rangle \left| 0\right\rangle _c \\ 
-\left| 211\right\rangle \left| 0\right\rangle _c
\end{array}
\stackrel{\text{Step(vii)}}{\longrightarrow } 
\begin{array}{c}
\left| 100\right\rangle \left| 0\right\rangle _c \\ 
\left| 101\right\rangle \left| 0\right\rangle _c \\ 
\left| 110\right\rangle \left| 0\right\rangle _c \\ 
-\left| 111\right\rangle \left| 0\right\rangle _c
\end{array}
,
\end{equation}
where $\left| ijk\right\rangle $ is abbreviation of the state $\left|
i\right\rangle _1\left| j\right\rangle _2\left| k\right\rangle _3$ of SQUIDs
($1,2,3$) with $i,j,k\in \{0,1,2\}$.

On the other hand, it is obvious that the following states of the system 
\begin{equation}
\;\left| 000\right\rangle \left| 0\right\rangle _c,\left| 001\right\rangle
\left| 0\right\rangle _c,\left| 010\right\rangle \left| 0\right\rangle
_c,\left| 011\right\rangle \left| 0\right\rangle _c
\end{equation}
remain unchanged during the operation. This is because: (a) the state $%
\left| 0\right\rangle $ of SQUID $1$ was not effected by the applied
microwave pulse, since the $\left| 0\right\rangle \leftrightarrow \left|
2\right\rangle $ transition and the $\left| 0\right\rangle \leftrightarrow
\left| 1\right\rangle $ transition of SQUID $1$ are far-off resonant with
the applied microwave pulse; and (b) no photon was emitted to the resonator
during the Step (ii) operation, when SQUID $1$ is initially in the state $%
\left| 0\right\rangle $. Hence, it can be concluded from Eq. (9) that the
three-qubit controlled phase gate (8) was achieved with three SQUIDs (i.e.,
the control SQUIDs $1$ and $2$, as well as the target SQUID $3$) after the
above process.

\begin{center}
{\bf B. }$n${\bf -qubit controlled phase gate}
\end{center}

A quantum controlled phase gate of $n$ qubits ($1,2,...,n$) is defined by
the following transformation 
\begin{equation}
\left| i_1i_2...i_n\right\rangle \rightarrow \left( -1\right) ^{i_1\times
i_2...\times i_n}\left| \text{ }i_1i_2...i_l...i_n\right\rangle ,
\end{equation}
where the subscript $l$ represents qubit $l$, $i_l\in \left\{ 0,1\right\} $,
and $\left| i_1i_2...i_l...i_n\right\rangle $ is a $n$-qubit computational
basis state. For $n$ qubits, there are a total number of $2^n$ computational
basis states, which form a set of complete orthogonal bases in a $2^n$%
-dimensional Hilbert space of the $n$ qubits. Eq. (11) shows that only when
the $n-1$ control qubits (the first $n-1$ qubits) are all in the state $%
\left| 1\right\rangle $, the state $\left| 1\right\rangle $ of the target
qubit (the last qubit) undergoes a phase flip, i.e., $\left|
11...1\right\rangle \rightarrow -\left| 11...1\right\rangle .$ While nothing
happens to all other $2^n-1$ computational basis states. In the following,
we will discuss how this gate can be achieved with $n$ SQUIDs coupled to a
resonator.

The $n$ SQUIDs are labeled by $1,2,...,$ and $n.$ The first $n-1$ SQUIDs ($%
1,2,...,n-1$) act as control qubits while the SQUID $n$ is the target qubit.
Suppose that the $n$ SQUIDs ($1,2,...,n$) are initially decoupled from the
resonator (initially in the vacuum state). Examining the above operations
for the three-qubit controlled phase gate carefully, we find that the $n$%
-qubit controlled phase gate (11) can be obtained with the resonator mode
returning to the original vacuum state, by the sequence of operators 
\begin{equation}
U=U_1^{+}\otimes \left( \prod_{l=n-1}^1U_{lr}\right) \otimes U_{nr}^2\otimes
\left( \prod_{l=1}^{n-1}U_{lr}\right) \otimes U_1,
\end{equation}
where $\prod_{i=1}^kU_{ir}\equiv U_{kr}U_{\left( k-1\right) r}\cdot \cdot
\cdot U_{2r}U_{1r};$ $U_1^{+}$ and $U_1$ denote the operators on SQUID $1$
with the following matrixes of 
\begin{equation}
U_1^{+}U_1=I,\;\text{ }U_1=\left( 
\begin{array}{cc}
0 & 1 \\ 
-1 & 0
\end{array}
\right)
\end{equation}
in the basis states $\left| 1\right\rangle _1=\left( 0,1\right) ^T$ and$%
\;\left| 2\right\rangle _1=\left( 1,0\right) ^T;$ $U_{lr}$ is the joint
operator on the SQUID $l$ and the resonator mode ($l=1,2,...,n-1$),
represented by the matrix 
\begin{equation}
U_{lr}=\left( 
\begin{array}{cc}
0 & -i \\ 
-i & 0
\end{array}
\right)
\end{equation}
in the basis states $\left| 0\right\rangle _l\left| 1\right\rangle _c=\left(
0,1\right) ^T$ and$\;\left| 2\right\rangle _l\left| 0\right\rangle _c=\left(
1,0\right) ^T;$ and $U_{nr}$ is the joint operator on the SQUID $n$ and the
resonator mode with the same matrix as the one described by (14) [in the
basis states $\left| 0\right\rangle _n\left| 1\right\rangle _c=\left(
0,1\right) ^T$ and$\;\left| 2\right\rangle _n\left| 0\right\rangle _c=\left(
1,0\right) ^T$].

From the description in the previous subsection, it can be seen that:

(i) $U_1$ ($U_1^{+}$) denotes the application of a $\pi $ microwave pulse ($%
\Omega _{12}\tau _{\mu w}=\pi ,$ where $\tau _{\mu w}$ is the pulse
duration) with $\phi =-\pi /2$ ($\pi /2$) and $\nu _{\mu w}=\nu _{12}$ (the $%
\left| 1\right\rangle \leftrightarrow \left| 2\right\rangle $ transition
frequency of SQUID $1$) to SQUID $1;$

(ii) $U_{lr}$ corresponds to the operation of bringing the $\left|
0\right\rangle \leftrightarrow \left| 2\right\rangle $ transition of SQUID $%
l $ ($l=1,2...,n-1$) to resonance with the resonator for an interaction time 
$\tau _l=\pi /\left( 2g_l\right) $ (i.e., a $\pi /2$ Jaynes-Commings pulse);
and

(iii) $U_{nr}^2$ indicates two $\pi /2$ Jaynes-Commings pulses resonant with
the $\left| 0\right\rangle \leftrightarrow \left| 2\right\rangle $
transition of SQUID $n,$ which are, when combined together, equivalent to a $%
\pi $ Jaynes-Commings pulse (i.e., $\tau _n=\pi /g_n$).

We remark that all other basis states of the system involved in each step of
the above operations, which form a complete set of orthonormal states
together with the basis states described above, are not affected by (a)
setting the microwave pulse to be largely detuned from the $\left|
0\right\rangle \leftrightarrow \left| 2\right\rangle $ transition and the $%
\left| 0\right\rangle \leftrightarrow \left| 1\right\rangle $ transition of
SQUID $1$ and (b) setting the $\left| 1\right\rangle \leftrightarrow \left|
2\right\rangle $ transition and the $\left| 0\right\rangle \leftrightarrow
\left| 1\right\rangle $ transition of each SQUID far-off resonant from the
resonator mode.

The method presented here for realizing the $n$-qubit controlled phase gate
(11) is the extension of the three-qubit version (8) described in the
previous subsection. This can be seen as follows. From Eq. (9), one can see
that the three-qubit controlled phase gate (8) was realized essentially
through the operation of step (iv). This operation lead to a phase shift on
the state of SQUID $3$ (i.e., $\left| 0\right\rangle \rightarrow -\left|
0\right\rangle $ and $\left| 1\right\rangle \rightarrow \left|
1\right\rangle $) with the aid of the photon, when the two control SQUIDs $1$
and $2$ are initially in the basis state $\left| 11\right\rangle .$ But,
when the two control SQUIDs $1$ and $2$ are initially in the other basis
states, this operation results in no change to the state of SQUID $3,$ due
to the fact that no photon was left in the resonator mode after the step
(iii). Similarly, the realization of the $n$-qubit controlled phase gate
(11) described above is mainly based on the operation described by $%
U_{nr}^2, $ which causes a phase shift on the state of SQUID $n$ (the
target) with the assistance of the photon when the $n-1$ control SQUIDs ($%
1,2,...,n-1$) are initially in the basis state $\left| 11...1\right\rangle .$
Note that only for this initial basis state, the photon originally created
by the operation $U_{1r}U_1$ would remain in the resonator mode after the
operation $\prod_{l=2}^{n-1}U_{lr}$. In addition, similar to the operations
of steps (v)-(vii) for the three-qubit controlled phase gate (8), the
operation described by $U_1^{+}\otimes \left( \prod_{l=n-1}^1U_{lr}\right) $
has the photon emitted back into the resonator mode from the system of
SQUIDs ($2,3,...n-1$) and finally absorbed by SQUID $1$. Hence, the
resonator mode returns to the original vacuum state and the $n$ SQUIDs ($%
1,2,...,n$) are back to the initial states except a phase flip to the $n$%
-qubit basis state $\left| 11...1\right\rangle $.

Before closing this section, some points may need to be addressed here.

(a). The irrelevant SQUIDs in each step of the operation need to be
decoupled from the resonator/pulse during the resonator/pulse-SQUID
interaction. The resonator mode needs to be not excited during the
application of the microwave pulse. In addition, the coupling of the levels $%
\left| 0\right\rangle $, $\left| 1\right\rangle ,$ and $\left|
2\right\rangle $ with the other levels should be negligible for each SQUID.
In principle, these conditions can be satisfied by adjusting the level
spacings of the SQUIDs. Note that for a SQUID, the level spacings can be
changed readily by varying the external flux $\Phi _x$ or the critical
current $I_c$ (e.g., for variable barrier rf SQUIDs) [50].

(b). It is not necessary to have a single-mode resonator since for a
multi-mode resonator one can choose one mode to interact with the SQUIDs
while have all other modes well decoupled from the three lowest levels of
the SQUIDs (e.g., with proper device parameters).

(c). The method presented here is applicable to a 1D, 2D, or 3D microwave
resonator/cavity as long as the conditions described above can be met.

(d). As is well known, a $n$-qubit controlled-NOT (CNOT) gate (known as the
Toffoli gate for $n=3$) can be obtained by combining the $n$-qubit
controlled phase gate (11) with two single-qubit Hadamard gates, which are
performed on the target qubit before and after the $n$-qubit controlled
phase gate (11) respectively (see Fig. 3). Each of the single-qubit Hadamard
transformations, $\left| 0\right\rangle \rightarrow \frac 1{\sqrt{2}}\left(
\left| 0\right\rangle +\left| 1\right\rangle \right) $ and $\left|
1\right\rangle \rightarrow \frac 1{\sqrt{2}}\left( \left| 0\right\rangle
-\left| 1\right\rangle \right) ,$ can be done with a $\pi /2$ microwave
pulse resonant with the $\left| 0\right\rangle \leftrightarrow \left|
1\right\rangle $ transition of the target SQUID qubit. When combining with
the above quantum controlled phase gate operations, one obtains the $n$%
-SQUID qubit CNOT gate.

(e). The present method provides a simple way to implement a multi-qubit
CNOT gate with SQUIDs coupled to a resonator. To see this, let us consider
the simple case of $n=3.$ It is known that construction of a Toffoli gate
requires at least {\it six} two-qubit CNOT gates and {\it ten} single-qubit
gates (i.e., two Hadamard, one phase, and seven $\pi /8$ gates) [51]. Note
that a two-qubit CNOT gate consists of a two-qubit controlled phase gate and
two single-qubit Hadamard gates as described above. Therefore, using the
conventional gate-constructing technique, at least 28 steps of operations
will be necessary to realize a Toffoli gate, assuming that the realization
of a single-qubit gate or a two-qubit controlled phase gate requires
one-step operation only. However, as discussed above, the present method
only needs 9 steps of operations. That is, seven steps of operations for the
three-qubit controlled phase gate (8) plus two steps of operations for the
two Hadamard gates. Finally, it is obvious that the simplicity of the
present method in constructing a $n$-qubit CNOT gate may become more
apparent with the increment of $n,$ when compared with the use of the
conventional gate-decomposition protocol.

\begin{center}
{\bf IV. DISCUSSION}
\end{center}

In this section we discuss issues that are important to experimental
implementation. Without loss of generality, let us consider performing a $n$%
-qubit controlled phase gate with $n$ identical SQUIDs ($1,2,...,n$) at
locations where the ${\bf B}_r$ fields are the same (e.g., antinodes of the
cavity field). In this case, we have $g_l=g$ ($l=1,2,...,n$). For the method
to work, the total operation time $\tau =2n\left( \tau _r+\tau _a\right)
+2\tau _{\mu w}$ [$\tau _r=\pi /\left( 2g\right) $] should be much shorter
than the energy relaxation time $\gamma _2^{-1}$ of the level $\left|
2\right\rangle ,$ and the lifetime of the resonator mode $\kappa
^{-1}=Q/2\pi \nu _r,$ where $Q$ is the (loaded)\ quality factor of the
resonator. Here, $\tau _a$ is the typical time required for adjusting the
level spacings of a single SQUID. In addition, direct coupling between
SQUIDs needs to be negligible since this interaction is not intended.

In principle, these requirements can be realized, since: (i) $\tau _r$ can
be reduced by increasing the coupling constant $g$, (ii) $\tau _a$ can be
shortened by rapid adjustment of the level spacings of the SQUIDs, (iii) $%
\kappa ^{-1}$ can be increased by employing a high-$Q$ resonator so that the
resonator dissipation is negligible during the operation, (iv) the SQUIDs
can be designed so that the energy relaxation time $\gamma _2^{-1}$ of the
level $\left| 2\right\rangle $ is sufficiently long, and (v) direct
interaction between SQUIDs is negligible as long as the following condition
can be met 
\begin{equation}
H_{s-s}\ll H_{s-r},H_{s-\mu w},
\end{equation}
where $H_{s-s}$ is the interaction energy between the two nearest SQUIDs, $%
H_{s-r}$ is the SQUID-resonator interaction energy, and the SQUID-microwave
interaction energy.

It is straightforward to see that the condition (15) can be realized if 
\begin{equation}
\zeta =\frac{ML_l^{-1}L_{l+1}^{-1}\max \left\{ \phi _{ij}^{(l)}\phi
_{ij}^{(l+1)}\right\} \Phi _0^2}{\min \left\{ \hbar g_l,\hbar \Omega
_{12}\right\} }\ll 1.
\end{equation}
Here, $M$ is the mutual inductance between the two nearest SQUIDs $l$ and $%
l+1$ ($l=1,2,...,n-1$), $\phi _{ij}^{\left( l\right) }$ ($\phi _{ij}^{(l+1)}$%
) $\equiv \left\langle i\right| \Phi \left| j\right\rangle /\Phi _0$ is the
magnetic dipole coupling matrix element between levels $\left|
i\right\rangle $ and $\left| j\right\rangle $ of SQUID $l$ ($l+1$), and $%
ij=01,02,$ or $12.$

For the sake of definitiveness, let us consider the experimental feasibility
of realizing a three-qubit controlled phase gate using SQUIDs with the
parameters listed in Table 1. Note that SQUIDs with these parameters are
readily available at the present time [22,23,52]. With the choice of these
parameters, the SQUIDs have the desired three-level structure as depicted in
Fig. 1. For a superconducting one dimensional standing-wave CPW (coplanar
waveguide) resonator with the parameters listed in Table 1 and SQUIDs placed
along the cavity axis (Fig. 4), one has $M_{sr}\sim 100$ pH. When each SQUID
is located at one of the antinotes of the resonator mode (Fig. 4), a simple
calculation shows $g\sim 7.5\times 10^9$ s$^{-1},$ resulting in $\tau
_r\approx 0.2$ ns. With the choice of $\tau _{\mu w}\sim \tau _a\sim \tau _r$%
, one has $\tau \sim 2.8$ ns, which is much smaller than $\gamma _2^{-1}\sim
3.2$ $\mu $s and $\kappa ^{-1}\sim 41.7$ ns for a resonator with $%
Q_r=3\times 10^3$. Note that superconducting CPW resonators with higher
quality factors have been demonstrated by recent cavity QED experiments with
superconducting charge qubits [8].

For a resonator with $\nu _r=11.4$ GHz, the wavelength of the resonator mode
is $\lambda \sim 10.5$ mm. When each SQUID is placed at an antinode of the $%
{\bf B}_r$ field (Fig. 4), one has $D\sim 5.3$ mm, where $D$ is the distance
between the two nearest SQUIDs. A simple estimate gives $M<0.1$ aH,
resulting in $\zeta <10^{-8}$ for the parameters considered above. Thus, the
condition of negligible direct coupling between SQUIDs is very well
satisfied.

The above analysis demonstrates that the realization of a three-qubit
controlled phase gate is possible using SQUIDs and a resonator. We remark
that a quantum controlled phase gate with a larger number of qubits can in
principle be obtained by increasing the length of the resonator though the
conditions of $\tau \ll \gamma _2^{-1},$ $\kappa ^{-1}$ becomes increasingly
difficult to satisfy.

We emphasize that the primary purpose of this work is to provide a new
approach to implement a multiqubit quantum controlled phase gate with
SQUIDs. However, we note that: (a) when coupled to a cavity mode, many
physical qubit systems (such as atoms, quantum dots, and superconducting
charge qubits) have the same type of qubit-cavity interaction described by
the Hamiltonian (2), and (b) the condition, i.e., the $\left| 0\right\rangle
\leftrightarrow \left| 2\right\rangle $ transition being resonant while the
transition between any other two levels being far-off resonant with the
resonator, can always be obtained via the adjustment of the level spacings
(e.g., for quantum dots and atoms, the level spacings can be changed via
adjusting the external electric field [53]). Therefore, it is
straightforward to show that the method can be generalized to realize the
multiqubit controlled phase gate in other types of qubit systems with the $%
\Lambda $-type three level configuration within cavity QED.

\begin{center}
{\bf V. CONCLUSION}
\end{center}

Before conclusion, we should point out that the idea of tuning the
individual qubits in and out of resonant from the cavity mode was previously
proposed for superconducting charge qubits [54]. Rather, our scheme is for a
different system and it differs in the details of both the qubits and the
coupling structure. In our case, we consider a system consisting of flux
qubits (SQUIDs) coupled to a microwave resonator/cavity, while the system
described in [54] comprises charge qubits and a LC-oscillator mode in the
circuit. In addition, the idea of performing a phase shift via the
assistance of the cavity photon was proposed earlier for realizing a {\it %
two-qubit} quantum controlled phase gate with trapped ions [46]. However, to
the best of our knowledge, our scheme is the first to demonstrate that a
quantum controlled phase gate{\it \ with a large number of qubits} can in
principle be achieved within cavity QED which is of great importance.

In summary, we have presented a method to realize a multiqubit controlled\
phase gate with SQUIDs coupled to a microwave resonator, which operates
essentially by exchanging a single photon between the control SQUIDs and the
resonator mode before and after a phase shift performed on the target SQUID.
The method has the following advantages. (i) Only one SQUID interacts with
the microwave pulse; (ii) No auxiliary SQUIDs or measurement is needed
during the entire operation, thus the hardware resources is significantly
reduced and the operation is greatly simplified; (iii) As tunneling between
the qubit levels $\left| 0\right\rangle $ and $\left| 1\right\rangle $ is
not required during the operation, decay from the level $\left|
1\right\rangle $ can be made negligibly small during the operation (via
prior adjustment of the potential barrier between the qubit levels $\left|
0\right\rangle $ and $\left| 1\right\rangle $ [50]) and therefore the
storage time of each qubit can be made much longer; (iv) The coupling
constants of SQUIDs with the resonator could be different, which makes the
present proposal much easier to implement since neither identical SQUIDs nor
exact placement of SQUIDs is needed; and (v) The method can in principle be
applied to obtain a $n$-qubit controlled phase gate with a large number $n$.
Finally, as discussed above, the present method is quite general, which can
be applied to implement a multiqubit controlled phase gate for other types
of physical qubit systems with the $\Lambda $-type three-level structure
within cavity QED.

\begin{center}
{\bf ACKNOWLEDGMENTS}
\end{center}

This work was partially supported by National Science Foundation QuBIC
program (ECS-0201995), ITR program (DMR-0325551), and AFOSR
(F49620-01-1-0439), funded under the Department of Defense University
Research Initiative on Nanotechnology (DURINT) Program and by the ARDA.

\begin{center}
{\large Figure and Table Captions\\}
\end{center}

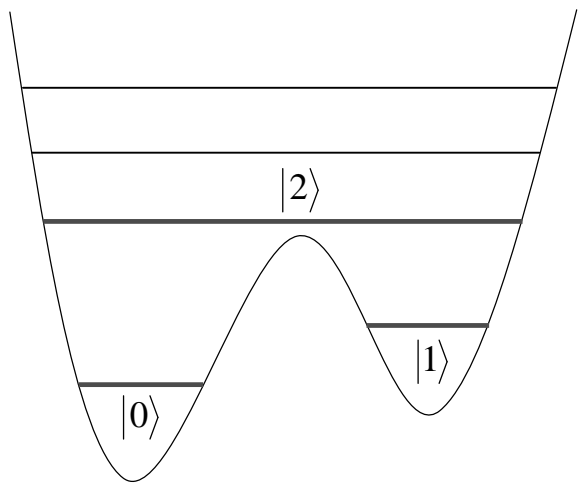
FIG. 1. Level diagram of a SQUID with the three lowest levels $\left|
0\right\rangle ,$ $\left| 1\right\rangle $ and $\left| 2\right\rangle $
forming a $\Lambda $-type structure.

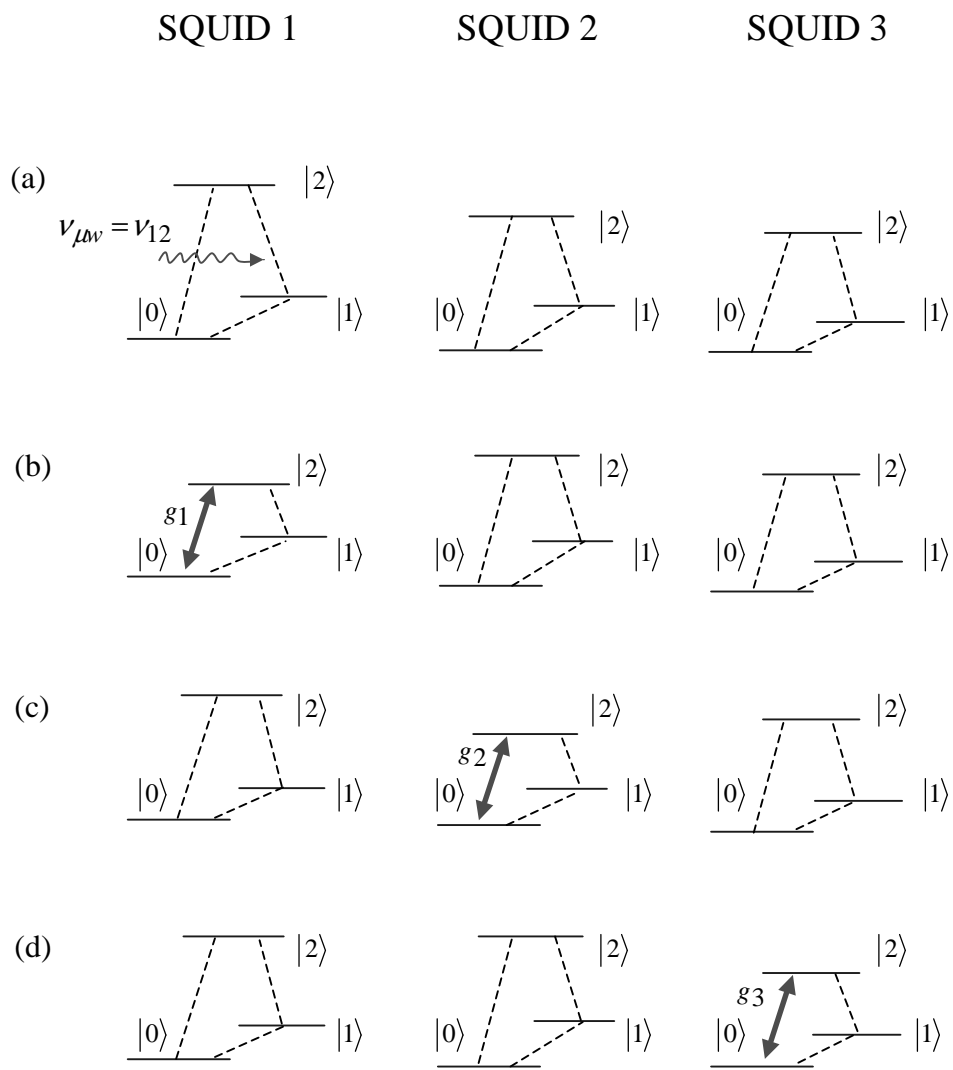
FIG. 2. Illustration for the change of the level structure (reduced) of
three SQUIDs ($1,2,3$) during a three-qubit controlled phase gate
performance. In (a), (b), (c), and (d), figures from left to right represent
the level structures for SQUIDs $1,2,$ and $3$, respectively; the
non-identical level spacings of the SQUIDs could be caused by nonuniform
device parameters. $g_1,$ $g_2,$ and $g_3$ are the resonant coupling
constants between the resonator mode and the $\left| 0\right\rangle
\leftrightarrow \left| 2\right\rangle $ transition of SQUIDs $1,2,$ and $3$,
respectively. The difference for $g_1,$ $g_2,$ and $g_3$ is due to device
parameter non-uniformity or non-exact placement of each SQUID. $\nu _{\mu w}$
is the frequency of the applied microwave pulse while $\nu _{12}$ is the $%
\left| 1\right\rangle \leftrightarrow \left| 2\right\rangle $ transition
frequency for SQUID $1$. In (a), the level spacings of SQUID $1$ is set to
be much different from that of SQUIDs $2$ and $3$, such that SQUIDs $2$ and $%
3$ are decoupled from the applied pulse. The transition between any two
levels linked by a dashed line is far-off resonant with the resonator mode.

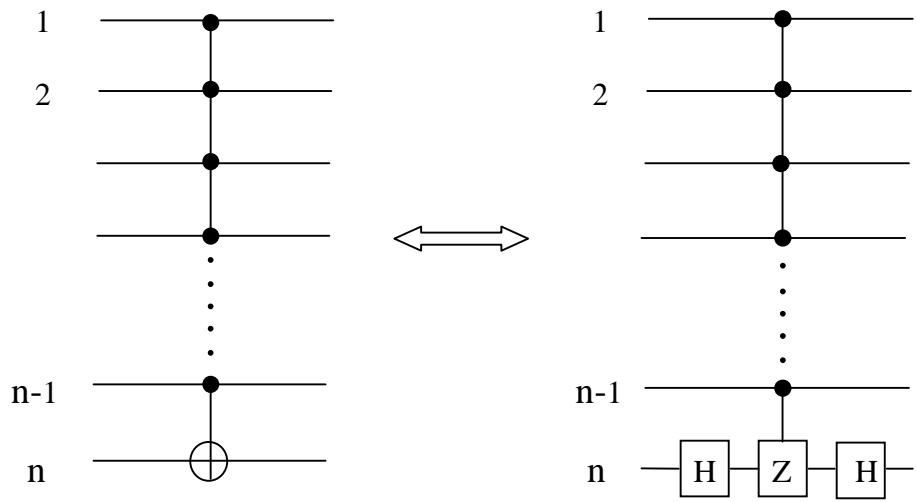
FIG. 3. Relationship between a $n$-qubit controlled NOT gate and a $n$-qubit
controlled phase gate. For the circuit on the left side, the element denoted
by $\oplus $ corresponds to a controlled NOT (with $n-1$ controls on the
filled circles); if the $n-1$ controls are all in the state $\left|
1\right\rangle ,$ then the state at $\oplus $ is bit-flipped. On the other
hand, for the circuit on the right side, the element Z represents a Pauli
rotation $\sigma _z,$ i.e., a phase flip operation (with $n-1$ controls on
the filled circles). Namely, if the $n-1$ control qubits are all in the
state $\left| 1\right\rangle ,$ then the state $\left| 1\right\rangle $ at Z
is phase-flipped as $\left| 1\right\rangle \rightarrow -\left|
1\right\rangle $ while nothing happens to the state $\left| 0\right\rangle $
at Z. In addition, the element containing H corresponds to a Hadamard
transformation described by $\left| 0\right\rangle \rightarrow \frac 1{\sqrt{%
2}}\left( \left| 0\right\rangle +\left| 1\right\rangle \right) $ while $%
\left| 1\right\rangle \rightarrow \frac 1{\sqrt{2}}\left( \left|
0\right\rangle -\left| 1\right\rangle \right) .$

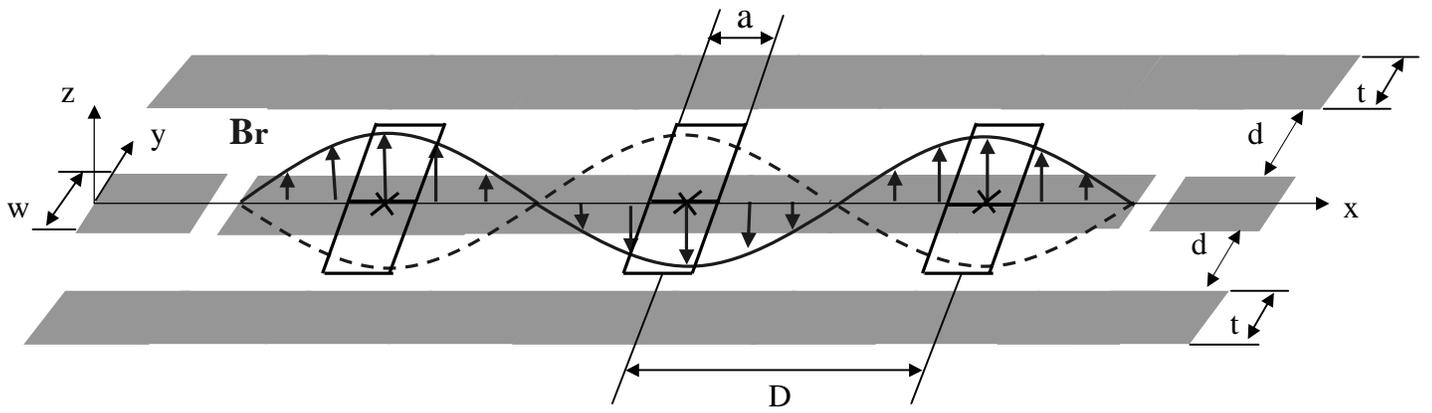
FIG. 4. Sketch of the setup for three SQUIDs ($1,2,3$) and a standing-wave
quasi-one dimensional CPW resonator (Not drawn to scale). Each SQUID is
placed in the plane of the resonator between the two lateral ground planes
(i.e., the $x$-$y$ plane) and at an antinode of the ${\bf B}_r$ field. The
two curved lines represent the standing-wave ${\bf B}_r$ field, which is in
the $z$-direction.

TABLE 1. Parameters for a SQUID-resonator. $\beta _L$ is the SQUID's
potential shape parameter, $R$ is the SQUID's effective damping resistance,
and $S$ is the surface bounded by the loop of the SQUID with width $a$ and
length $b$. $\gamma _2^{-1}$ ($\gamma _1^{-1}$) is the energy relaxation
time of the level $\left| 2\right\rangle $ ($\left| 1\right\rangle $). $\nu
_{02}$ ($\nu _{12}$) is the $\left| 0\right\rangle \leftrightarrow \left|
2\right\rangle $ ($\left| 1\right\rangle \leftrightarrow \left|
2\right\rangle $) transition frequency. $\phi _{ij}\equiv \left\langle
i\right| \Phi \left| j\right\rangle /\Phi _0$ is the magnetic dipole
coupling matrix element between levels $\left| i\right\rangle $ and $\left|
j\right\rangle $ ($i=0,$ $1;$ $j=2$). $l$ is the length of the quasi-one
dimensional CPW resonator, $\lambda $ is the wavelength of the resonator
mode with frequency $\nu _r,$ $d$ is the gap between the center conductor
and the adjacent ground plane, $w$ is the width of the center conductor, $t$
is the width of each ground plane, $L_0$ is the inductance per unit length
of the waveguide, and $\varepsilon _e$ is the effective relative dielectric
constant.

\end{document}